# High visibility two-photon interference of frequency-time entangled photons generated in a quasi-phase-matched AlGaAs waveguide

Peyman Sarrafi[1], Eric Y. Zhu[1], B. M. Holmes[2], D.C. Hutchings[2], Stewart Aitchison[1], Li Qian[1]

[1]*Dept. of Electrical and Computer Engineering, Univ. of Toronto, 10 King's College Road, Toronto, Ontario, Canada M5S 3G4*

[2]*School of Engineering, University of Glasgow, Glasgow G12 8QQ, Scotland, U.K.*

*peyman.sarrafi@utoronto.ca*

**Abstract:** We demonstrate experimentally the frequency-time entanglement of photon pairs produced in a CW-pumped quasi-phased-matched AlGaAs superlattice waveguide. A visibility of $96.0 \pm 0.7\%$ without background subtraction has been achieved, which corresponds the violation of Bell inequality by 52 standard deviations.

An entangled Photon Pair Source (PPS) is an important component in quantum communication and quantum cryptography [1-3] where quantum entanglement is distributed over long distances. A number of on-chip sources of photon pairs [4, 5] have been investigated with the aim of replacing conventional bulk sources. III-V direct-bandgap semiconductor-based photon pair sources [6-8] have recently attracted attention as such platforms allow for a monolithically integrated entangled photon source with a pump laser fabricated on the same substrate [9]. Specifically, the (Al)GaAs platform for a PPS at telecom wavelengths (~1550 nm) has a number of inherent advantages over alternative approaches. First, there exists a mature fabrication technology for this platform. Second, (Al)GaAs possesses a relatively large second-order optical nonlinearity, $\chi^{(2)}$, which facilitates the Spontaneous Parametric Down-Conversion (SPDC) process, commonly used for generating the entangled photon pairs. Through this process, a pump photon is annihilated and two photons are created that meet the criteria of energy conservation and phase-matching. An inherent advantage of the SPDC process, as opposed to the spontaneous four-wave-mixing ($\chi^{(3)}$) process, is that photon pairs are created spectrally far from the pump photons, therefore, pump and Raman noise can be easily, and nearly completely, filtered. Third, the direct-bandgap of AlGaAs allows for pump emission around 775 nm, leading to generated pairs at ~1550 nm. However, the phase matching requirement is arguably the most challenging issue in this material system since AlGaAs does not possess a natural material birefringence, and hence other phase matching approaches must be used. We have previously developed a quasi-phase-matching (QPM) technique based upon quantum-well intermixing in an AlGaAs superlattice waveguide [10]. Unlike most other phase-matching techniques investigated in III-V semiconductors [6-7], our method allows the QPM pattern to be defined by lithography post-growth, and can, therefore, provide a range of phase-matching wavelengths across a number of devices on the same chip. In addition, in contrast to the Bragg reflection waveguide structures of Ref. [7], our conventional waveguide design leads to high coupling efficiency and better mode shape, and facilitates the potential monolithic integration with conventional edge-emitting laser structures.

We have previously reported a CW-pumped correlated photon source in QPM AlGaAs superlattice waveguides [8]. The source has been shown to have very low noise (i.e. less than 1 percent of coincidences were accidental), and the pair production rate has been shown to be high, but the quantum properties of generated entangled photons had not yet been characterized. In this work, the time entanglement property of the down-converted photons is demonstrated using a Franson interferometer.

The PPS is made of a waveguide with a 0.6-µm-thick core layer of 14:14 monolayer GaAs/$Al_{0.85}Ga_{0.15}$As superlattice, sandwiched between 300nm $Al_{0.56}Ga_{0.44}As$ buffer layers, and 800 nm cladding layers of $Al_{0.60}Ga_{0.40}As$. There is an additional 1-μm-thick layer of $Al_{0.85}Ga_{0.15}As$ below the cladding to separate the optical mode from GaAs substrate. The waveguide mode profiles at both pump and signal/idler wavelengths (Fig.1) are well-confined and easy to couple into fiber or other waveguides.

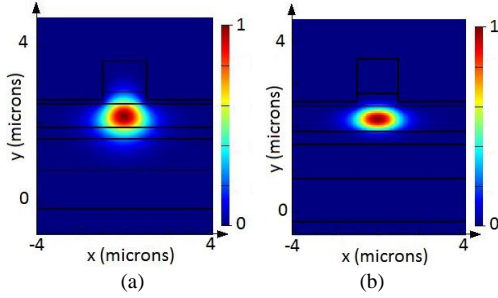

Fig. 1. (a) Simulated TE mode profile at 1550 nm (b) Simulated TM mode profile at 775 nm

We use ion implantation induced quantum well intermixing to define the QPM structure, which in this case has a period of 3.5 μm. Further details of the device fabrication can be found in reference [9]. Compared to the previous device we reported [8], here we use a more tightly confined waveguide (2 μm width vs. 3 μm) and an Au ion implantation mask duty cycle of 50 percent instead of 60 percent. As a result, we achieved a 5-fold enhancement in brightness ($8\times10^6$ pairs/s in a 8-nm bandwidth) for the same pump power (10mW).

The schematic of our experimental setup is shown in Fig. 2. The pump is a 773 nm CW laser (Toptica DL pro) with linewidth of smaller than 1 MHz, and is launched into the waveguide at TM polarization.

The waveguide output, containing type I down-converted photon pairs, is sent through a polarizer (passing TE) and a pump-suppression filter. The pump suppression filter can be a fiber based filter or more simply a 1 mm-thick layer of GaAs in free space. After pump filtering, fiber-based band-pass filters (BPF) [4, 5] are then used to deterministically separate the photon pairs into two 16-nm spectral bands, centered on 1570 nm and 1530 nm. (Note the choice of the BPFs is limited by availability, and the BPFs are not exactly symmetric with respect to the degenerate wavelength of 1546.0 nm.) As a result, the portion of the filter passbands that are frequency conjugate reduce the effective bandwidths to 8-nm (coherence time of ~1 ps).

Each of the BPFs sends its output to a commercial unbalanced silica Planar Lightwave Circuit (PLC) Mach Zehnder Interferometer (MZI). The two MZIs form a Franson interferometer which is used to characterize the entangled nature of the photon pairs. The delay difference between the short and long arms in both MZIs is 500 ps (2 GHz). Each of the MZIs is equipped with a Peltier cell to adjust and stabilize the overall device temperature. These Peltier cells have been used to match the time delay differences between the MZIs to an accuracy of better than 50 fs. This difference has been measured in a separate setup using white light interferometry and utilizing a third reference MZI with a tunable delay on one arm. Additionally, on the long arm of each MZI, a local heater is used to accurately control the phase difference over a range of 0-360°.

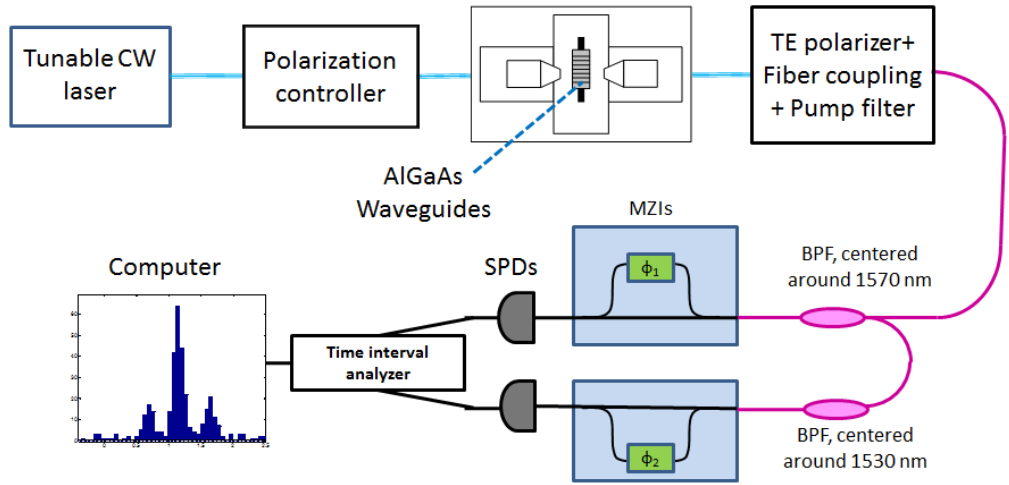

Fig. 2 Schematic of the setup for Franson interferometry used to characterize the AlGaAs-waveguide-based photon pair source.

Characterization of these heaters shows that the relative phase is proportional to the square of the applied voltage, as expected. The output of these two MZIs are connected to two, free-running, single photon detectors (SPDs, IDQ id220), with measured quantum efficiencies of 20% at 1550 nm. The SPD output electrical signals are sent to a commercial time interval analyzer (TIA, Picoquant Hydraharp) to record the difference in photon arrival time. The TIA has a measurement resolution of 64 ps which acts as a narrow time domain filter [11].

The total estimated loss associated with the waveguide-to-free-space coupling, TE polarizer, free- space-to-fiber coupling, BPF and MZI is ~20 dB (each branch) at 1550 nm. There is also a ~5 dB pump-to-waveguide coupling loss. The total length of fibers for each branch is less than 8 metres, thus the effects of dispersion can be neglected. Additionally, Franson interferometry requirements are satisfied [12], as the pump linewidth (1 MHz) is smaller than the free-spectral range of the MZIs (2 GHz), which is smaller than the spectral extent of the signal or idler photon (~10 nm at 1550 nm, or 1.3 THz).

In Fig. 3, two typical histograms show the coincidence measurements recorded by the TIA, corresponding to two different sets of voltages applied to the MZI heaters. For the data in Fig. 3 (a), the applied phase differences to long arms of MZI 1 and 2 are 270 and 180 degrees, respectively. For the data in Fig. 3 (b), the applied phase differences to long arms of MZI 1 and 2 are 90 and 180 degrees, respectively.

As shown in Fig. 3, the recorded histogram has three coincidence peaks. The left and right peaks correspond to a state when one photon goes through the short arm in one MZI and the other through the long arm in the other MZI. The middle peak results from the interference between the state where both photons traverse the short arms and the state where both traverse the long arms of MZIs.

This peak can be interpreted as the result of fourth order quantum interference of the following post-selected state [12]:

$$|\psi\rangle = \frac{1}{\sqrt{2}}\big[|short\rangle_i|short\rangle_s + e^{-i(\varphi_1+\varphi_2)}|long\rangle_i|long\rangle_s\big] \quad (1)$$

For photons from a perfect Einstein-Podolsky-Rosen (EPR) source, the total number of coincidences in the middle peak exhibits an interference fringe of unity visibility when the relative phases of the two MZIs are varied [12].

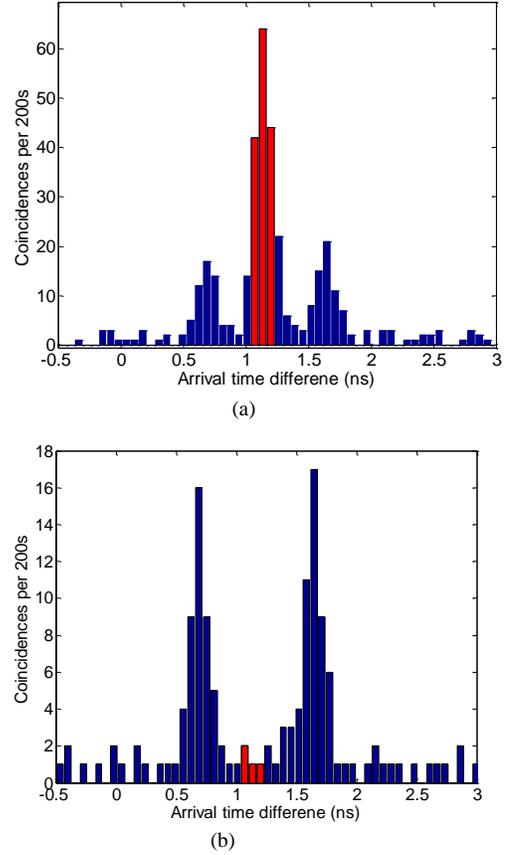

Fig. 3. The result of coincidence measurements at (a) $\phi_1 = 270°, \phi_2 = 180°,$ and (b) $\phi_1 = 90°, \phi_2 = 180°$ The red bins are used to calculate the coincidences for Fig. 4.

The two phases can be chosen to maximize (Fig. 3a) or minimize (Fig. 3b) the middle peak height.

A 212 ps electronic jitter, dominated by the 150 ps jitter of each SPD, limits the measurement accuracy of the arrival time differences between photons pairs. In Fig. 4(a), the coincidence counts corresponding to the middle peak, as indicated by the 3 red 64 ps time-bins in Fig. 3, have been plotted as a function of $\phi_1$ for set values of $\phi_2$ of 0, 90 and 180 degrees. Choosing three bins is a result of compromise between maximizing the coincidence counts of the central peak and minimizing the infiltration of counts from the adjacent peaks. The total width of the 3 bins is also equal to total electronic jitter.

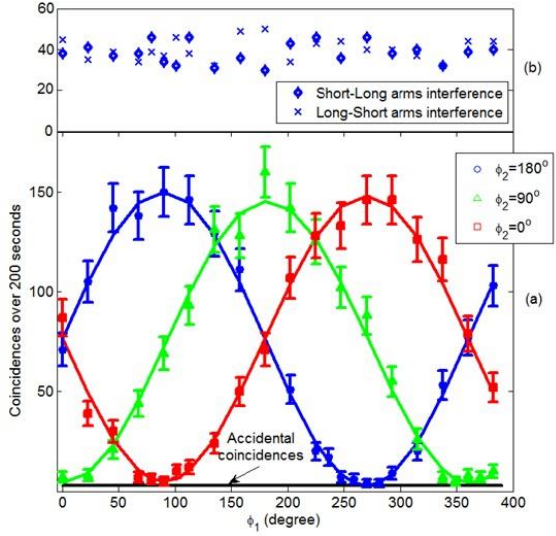

Fig. 4. Two photon interference measurements for three phase settings on the second MZI.(a) Middle peak coincidences shown in Fig. 3 plotted as a function of $\phi_1$ for three fixed $\phi_2$ values. (b) Left and right peak coincidences shown in Fig. 3 plotted as a function of $\phi_1$ for $\phi_2$ =180°

Sinusoidal visibilities, $\mathcal{V}(\phi_2)$, of 96.0±0.7% for $\phi_2$=180°, 94.3±0.7% for $\phi_2$=90° and 94.0±0.7% for $\phi_2$=0° are obtained. We expect the visibilities for all the $\phi_2$ angles settings to be the same, as the two-photon interference is a function of $\phi_1 + \phi_2$. However, there might be small measurement uncertainties associated with the voltage-phase relationship we used for the MZIs, which might have contribute to the difference in visibilities. The observed difference from 100% visibility is mainly due to the accidental coincidence (shown as a black line in Fig. 4), and would increase to average of ~99% with the subtraction of a constant accidental background. Note here the contributions to the noise include dark counts ($2\times10^3$ per second per detector) and background fluorescence ($8\times10^3$ per second per detector). The coincidence counts corresponding to the left (short-long arm interference) and right (long-short arm interference) peaks in Fig. 3 remained essentially unchanged during the experiment (Fig. 4(b)), verifying that the sinusoidal variation seen in the middle peak (Fig. 4(a)) is due to quantum interference. A Clauser-Horne-Shimony-Holt (CHSH)-Bell inequality [13] of S=2.687±0.013 is obtained from the raw visibilities, $\mathcal{V}(\phi_2)$, $S = \sqrt{2}(\mathcal{V}(90^o) + \mathcal{V}(180^o))$ [14]. It demonstrates the violation of Bell inequality by more than 52 standard deviations, confirming that down-converted photon pairs produced by the quasi-phase-matched AlGaAs superlattice waveguides are time-energy entangled.

In summary, two photon interference measurements obtained from a Franson interferometer have been used to characterize a CW-pumped, QPM, AlGaAs superlattice entangled photon pair source. The high raw visibility of approximately 95% (without background subtraction) indicates the high-purity, low-noise feature of this source, which has not been observed to date in any other types of AlGaAs-based photon pair sources. Together with its high brightness ($8\times10^6$ pairs/s) achieved with 10 mW pump power, its ability to be integrated planarly with a pump source, and its well-behaved modal profiles for fiber or waveguide coupling, our source proves a practical solution for future on-chip quantum communication circuits.

**References**
1. W. Tittel, J. Brendel, H. Zbinden, and N. Gisin, Physical Review Letters **84**, 4737 (2000)
2. J. Brendel, N. Gisin, W. Tittel, and H. Zbinden, Physical Review Letters **82**, 2594 (1999)
3. N. Gisin and R. Thew, Nat. Photonics **1**, 165 (2007).
4. N. Matsuda, H. L. Jeannic, H. Fukuda, T. Tsuchizawa, W. J. Munro, K. Shimizu, K. Yamada, Y. Tokura, and H. Takesue, Scientific report **2**, 817 (2012)
5. J. Sharping, K. Lee, M. Foster, A. Turner, B. Schmidt, M. Lipson, A. Gaeta, and P. Kumar, Opt. Express 14, 12388-12393 (2006)
6. R. T. Horn, P. Kolenderski, D. Kang, P. Abolghasem, C. Scarcella, A. D. Frera, A. Tosi, L. G. Helt, S. V. Zhukovsky, J. E. Sipe, G. Weihs, A. S. Helmy and T. Jennewein, Scientific reports, 3, 2314 (2013)
7. A. Orieux, A. Eckstein, A. Lemaître, P. Filloux, I. Favero, G. Leo, T. Coudreau, A. Keller, P. Milman, and S. Ducci., Physical Review Letters **110**, 16 (2013)
8. P. Sarrafi, E. Y. Zhu, K. Dolgaleva, B. M. Holmes, D. C. Hutchings, J. S. Aitchison, and L. Qian, Applied Physics Letters, **103**, 25 (2013)
9. F. Boitier, A. Orieux, C. Autebert, A. Lemaître, E. Galopin, C. Manquest, C. Sirtori, I. Favero, G. Leo, and S. Ducci, Physical Review Letters **112**, 183901 (2014)
10. S. J. Wagner, B. M. Holmes, U. Younis, I. Sigal, A. S. Helmy, J. S. Aitchison, and D. C. Hutchings, IEEE J. Quantum Electron. **47**, 834–840 (2011)
11. S. Dong, Q. Zhou, W. Zhang, Y. He, W. Zhang, L. You, Y. Huang, and J. Peng, Opt. Express **22**, 359-368 (2014)
12. J. D. Franson, Physical Review Letters **62**, 19 (1989)

13. J.F. Clauser, M.A. Horne, A. Shimony, R.A. Holt, Phys. Rev. Lett. **23**, 15 (1969)

14. E. Y. Zhu, Z. Tang, L. Qian, L. G. Helt, M. Liscidini, J. E. Sipe, C. Corbari, A. Canagasabey, M. Ibsen, and P. G. Kazansky, Phys. Rev. Lett. **108**, 213902 (2012)